\documentclass[10pts,showpacs,preprint,aps]{revtex4}
\usepackage{graphicx}
\usepackage{dcolumn}
\usepackage{bm}
\usepackage{amssymb}
\usepackage{amsmath}
\begin{document}
\setcounter{page}{1}
\vskip 2cm
\title
{The Quantum Yang Baxter conditions: The fundamental relations behind the Nambu-Goldstone theorem}
\author
{Ivan Arraut$^{(1,2)}$}
\affiliation{$^1$ The Open University of Hong Kong, 30 Good Shepherd Street, Homantin, Kowloon}
\affiliation{$^2$ Department of Physics, Faculty of Science, Tokyo University of Science,
1-3, Kagurazaka, Shinjuku-ku, Tokyo 162-8601, Japan}

\begin{abstract}
We demonstrate that when there is spontaneous symmetry breaking in any system, relativistic or non-relativistic, the dynamic of the Nambu-Goldstone bosons is governed by the Quantum Yang-Baxter equations. These equations describe the triangular dynamical relations between pairs of Nambu-Goldstone bosons and the degenerate vacuum. We then formulate a theorem and a corollary showing that these relations guarantee the appropriate dispersion relation and the appropriate counting for the Nambu-Goldstone bosons.                                
\end{abstract}
\maketitle 
\section{Introduction}
The Spontaneous symmetry breaking (SSB) phenomena was formulated by Nambu \cite{Nambu3}, who inspired by the theory of superconductivity proposed by Bardeen et. al \cite{Bardeen}, started to explore the apparent violations of gauge invariance on the model. A related problem was the formulation of quasiparticles inside the BCS theory done by Valantin and Bogoliubov \cite{Vala, Bolo}. In such a case, the quasiparticles did not carry a definite charge since they were a combination of electron and hole \cite{Nambu1, Nambu2}. Nambu then found that the problem is solved after introducing mass-less collective modes, today known as Nambu-Goldstone (NG) bosons. Then the principle of SSB emerged as a universal phenomena. It was applied for first time in particle physics for explaining the chiral symmetry breaking and the mass generation for nucleons \cite{Nambu3, Nambu4}. It helped in the formulation of the Higgs mechanism in particle physics \cite{Nambu5}, which provided the possibility of doing the electroweak unification \cite{Salam}. The original formulation of the NG theorem says that the number of Nambu-Goldstone bosons ($N_{NG}$) is equal to the number of broken generators ($N_{BG}$). In addition, in ordinary situations, the Nambu-Goldstone bosons have a linear dispersion relation, which is the case for mass-less particles. In some circumstances however, there is a mismatch between $N_{NG}$ and $N_{BG}$ as well as an unusual behavior for the dispersion relations for the Nambu-Goldstone bosons \cite{Murayama1, Murayama2, Murayama}. The problem of counting Nambu-Goldstone bosons as well as the understanding of the dispersion relations is not new and has been analyzed in the past. A solution was suggested by Nielsen and Chadha in \cite{Murayama1} and Nambu himself had already done some formulations explaining the apparent discrepancy between number of broken symmetries of the system and the number of Nambu-Goldstone bosons \cite{Murayama2}. One interesting formulation was the one proposed by Watanabe, Brauner and Murayama in \cite{Murayama}. Analysis of the Higgs mechanism under these special circumstances was done in \cite{Murayama5}. Other interesting approximations to this problem have been done in \cite{Murayama6}, where a relation between the Nambu-Goldstone boson and the order parameter was proposed. All these previous approaches were able to explain some rules for the counting of Nambu-Goldstone bosons as well as for the dispersion relations. Some of the explanations related to this behavior were centered in the redundancies of the broken generators or on the order of the derivative terms in the effective action \cite{Murayama}. Fundamental relations, explaining in a simplified way the dynamic of the Nambu-Goldstone bosons in different situations where the discrepancies appear were not formulated. In this paper, we find that surprisingly the Quantum Yang-Baxter equations (QYBE), in the same standard way as they are formulated in \cite{integrability}, describe the appropriate dynamic of pairs of Nambu-Goldstone bosons. This surprising and unexpected connection can be summarized in a theorem and a corollary based on the number of independent histories (Yang-Baxter triangles) describing the interaction of a pair of Nambu-Goldstone bosons. In the most general situations, if we have a pair of Nambu-Goldstone bosons ($n$ and $n'$) interacting along some region of the degenerate vacuum, we can define two independent histories related to each other under the exchange of particles ($n\to n'$). If the two histories are equivalent, having then only one independent history, then $n=n'$ and we have a single degree of freedom with quadratic dispersion relation. If the two histories remain independent, then we have two degrees of freedom (one for each Nambu-Goldstone boson) and the dispersion relation is linear. Surprisingly, the QYBE generate the necessary constraints which help us to obtain in a natural way the correct dispersion relations. The theorem described in this paper is formulated in terms of operators ($R$-matrices), suggesting that if the product of three matrices formed by {\bf1)}. The Nambu-Goldstone field. {\bf 2).} The pair of matrices corresponding to the broken generators; satisfy the QYBE after summing over the degenerate vacuum, then the appropriate dispersion relation for the Nambu-Goldstone bosons is obtained naturally. The proposed triangular relation, based on the sum of histories, solves automatically the problem of counting ($N_{NG}\neq N_{BG}$). The ordinary approach, suggests that when the symmetry is spontaneously broken, one vacuum among many is selected arbitrarily due to small fluctuations in the thermodynamic limit. On the other hand, in the formulation done in this paper, the application of the QYBE, requires the sum over all the possible vacuums. Here we make the Mathematical justification about the sum carried out over the degenerate vacuum, even if each vacuum represents in principle a different Hilbert space in the thermodynamic limit. Besides this, we will have as many different vacuums as independent broken generators are. The reason for this coincidence is the fact that the action of a broken generator over any vacuum is to rotate (change) it toward a different one. In different vacuums we will have different vacuum expectation values of the order parameter. The bi-linear character of the $R$-matrices makes the arguments formulated in this paper ideal for them to be adapted to physical systems involving quark-antiquark and quark-quark interactions as the ones analyzed in \cite{Murayama3}. In the same way, the formulation done in this paper is relevant for situations where we have a bi-linear object as an order parameter. This is the case of the chiral condensation analyzed in \cite{Murayama4, Field1}. The paper is organized as follows: In Sec. (\ref{eq:sec1}), we explain the spontaneous symmetry breaking phenomena by using the linear $\sigma$-model. In Sec. (\ref{eq:Sec2}), we explain the Quantum Yang-Baxter Equations (QYBE's) and we explain the relation between the broken generators and the order parameter based on the same relations. In Sec. (\ref{extension}), we analyze in detail the connection between the spontaneous symmetry breaking condition and the QYBE. In Sec. (\ref{The counting}) we briefly analyze how we can infer the number of Nambu-Goldstone bosons in a system by just looking at the rank of the $R$-matrices. In Sec. (\ref{Hiddenla?}), we formulate a theorem with its corollary connecting the $R$-matrices with the number of Nambu-Goldstone bosons and the dispersion relations. We also justify in this section the sum over the degenerate vacuum which makes it valid the Yang-Baxter formulation. Finally in Sec. (\ref{Conclusions}), we conclude.                

\section{Spontaneous symmetry breaking: The linear $\sigma$ model}   \label{eq:sec1}

Here we consider the the linear $\sigma$-model as an example of a system where the spontaneous symmetry breaking phenomena appears. Having $N$ scalar fields defined by $\phi^i(x)$, we can define a Lagrangian

\begin{equation}   \label{eq:Thelag}
\pounds=\frac{1}{2}(\partial_\mu\phi^i)^2+\frac{1}{2}\mu^2(\phi^i)^2-\frac{\lambda}{4}[(\phi^i)^2]^2,
\end{equation}
which is invariant under the following transformation

\begin{equation}
\phi^i\to R^{ij}\phi^j.
\end{equation}
This transformation is just a representation of the $O(N)$ group, namely, the group of orthogonal matrices in $N$ dimensions. The potential of the Lagrangian (\ref{eq:Thelag}) is given by

\begin{equation}
V(\phi^i)=-\frac{1}{2}\mu^2(\phi^i)^2+\frac{\lambda}{4}[(\phi^i)^2]^2.
\end{equation}    
This potential has a minimum when

\begin{equation}   \label{eq:Iseelanola}
(\phi^i_0)^2=\frac{\mu^2}{\lambda}.
\end{equation}
From this previous condition, we can determine the magnitude of $\phi^i_0$, but not its direction. Then the direction is in principle arbitrary. We can select some arbitrary direction, for example we can select $\phi^i_0$ as

\begin{equation}
\phi^i_0=(0,0,...,0,v),
\end{equation} 
with $v=\mu/\sqrt{\lambda}$. If we re-define the vacuum in agreement with the shift

\begin{equation}
\phi(x)=(\pi^k(x), v+\sigma(x)),
\end{equation}
with $k=1,...,N-1$. The Lagrangian in terms of the fields $\pi^k(x)$ ans $\sigma(x)$ becomes

\begin{equation}
\pounds=\frac{1}{2}(\partial_\mu\pi^k)^2+\frac{1}{2}(\partial_\mu\sigma)^2-\frac{1}{2}(2\mu^2)\sigma^2-\sqrt{\lambda}\mu\sigma^3-\sqrt{\lambda}\mu(\pi^k)^2\sigma-\frac{\lambda}{4}\sigma^4-\frac{\lambda}{2}(\pi^k)^2\sigma^2-\frac{\lambda}{4}[(\pi^k)^2]^2.
\end{equation}
This Lagrangian clearly contains $N-1$ massless $\pi^k$-fields and one massive field $\sigma$. Then there are $N-1$ broken generators and they correspond to the Nambu-Goldstone bosons in agreement with the standard formulation. Here we will explain how this results comes out. 

\begin{figure}
	\centering
		\includegraphics[width=10cm, height=8cm]{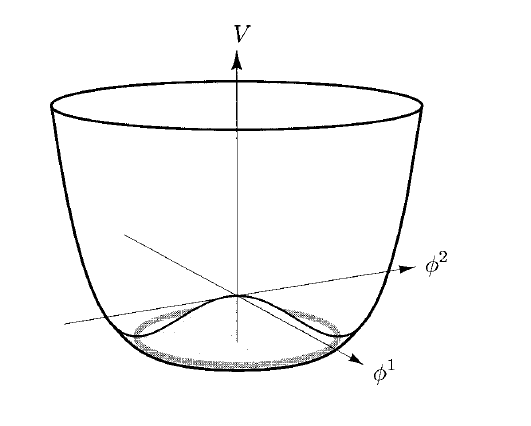}
	\caption{The potential for the spontaneous breaking of symmetry of the $O(N)$ symmetry, for the case of $N=2$. Taken from \cite{Field1}.}
	\label{fig:momoko}
\end{figure} 

\subsection{The Nambu-Goldstone theorem: Standard counting rule}

We have seen in the previous explanations that the $\sigma$-model provides a basic example of spontaneous symmetry breaking where the symmetry $O(N)$ is broken and then the symmetry $O(N-1)$ remains after selecting some arbitrary vacuum. In general, the previously defined potential $V(\phi^i)$, can be expanded around the minimum as

\begin{equation}   \label{eq:invariant}
V(\phi)=V(\phi_0)+\frac{1}{2}(\phi-\phi_0)^a(\phi-\phi_0)^b\left(\frac{\partial^2}{\partial\phi^a\partial\phi^b}V\right)_{\phi_0}+....,
\end{equation}  
where $\partial V/\partial\phi^a=0$. The mass matrix for the modes is symmetric and given by the coefficient

\begin{equation}
\left(\frac{\partial^2}{\partial\phi^a\partial\phi^b}V\right)_{\phi_0}=m^2_{ab}\geq0.
\end{equation}
This previous condition is a consequence of the fact that $\phi_0$ represents a minimum. At this point we will assume that the full action given by

\begin{equation}
\pounds=(kinetic\;\; terms)-V(\phi),
\end{equation} 
is invariant under the the action of the group $G=O(N)$. In addition, we assume that the selected vacuum state is invariant under the action of a subgroup of $G$, given by $H=O(N-1)$. In such a case, in general circumstances, the vacuum state is not invariant under the action of the full group $G$. In summary, we have the following conditions

\begin{equation}
G:\;\;\phi_0 ^{a'} =U(g)\phi_0 ^a\neq\phi_0^a,
\end{equation} 

\begin{equation}
H:\;\;\phi_0^{a'}=U(h)\phi_0 ^a=\phi_0^a,
\end{equation}
where $U(g)$ and $U(h)$ denote the representations of the groups $G$ and $H$  respectively. However, the potential $V(\phi)$ is still invariant under the action of the full group $G$. The action of this group on the potential expansion (\ref{eq:invariant}) gives the result

\begin{equation}   \label{eq:Globalsymmetry}
T^a(\phi)\frac{\partial}{\partial\phi^a}V(\phi)=0,
\end{equation}
where $T^a(\phi)$ is the generator of the group transformation. In the $U(g)$ representation for example, it would take the form

\begin{equation}
U(g)=e^{T^a\alpha}\approx \hat{I}+\alpha T^a\to U(h)\phi_0^a=\phi_0^a+\alpha T^a(\phi_0),
\end{equation}
where $T^a(\phi)$ denotes the action of the operator $T^a$ on the function state $\phi$. If we introduce the result (\ref{eq:Globalsymmetry}) inside the expansion (\ref{eq:invariant}), then we get the condition

\begin{equation}   \label{eq:Globalsymmetry2}
T^a(\phi_0)T^b(\phi_0)\frac{\partial^2}{\partial\phi^a\partial\phi^b}V(\phi)=T^a(\phi_0)T^b(\phi_0)m^2_{ab}=0.
\end{equation}
Note that if $T^a(\phi_0)=0$, namely, when the vacuum state selected is invariant under the action of the the group, then the corresponding mass component $m_{ab}$ is not necessarily zero. On the other hand, when the symmetry generator is broken, namely, when the group element belonging to $G$ does not leave the vacuum invariant, then $T^a(\phi_0)\neq0$ and then the mass components related to this condition are necessarily zero ($m_{ab}=0$). Then the number of broken generators are clearly related to the existence of gapless particles. In standard conditions, and for internal symmetries, the number of broken generators is equivalent to the number of Nambu-Goldstone bosons with linear dispersion relation \cite{Field1}. A linear dispersion relation is what we expect from massless particles. 

\subsection{Charge conservation}

We can define the conserved charge from the Lagrangian $\pounds$ as

\begin{equation}
j^a_\mu(x)=\frac{\partial\pounds}{\partial(\partial^\mu\phi)}\frac{\delta\phi(x)}{\delta\alpha^a},
\end{equation}
where $\delta\phi(x)/\delta\alpha$ corresponds to the field variations under symmetry transformations of the Lagrangian. The previously defined current is divergence-less and the corresponding charges are given by

\begin{equation}   \label{eq:Chargedef}
Q^a(x)=\int d^3xj^a_0(x).
\end{equation}  
In the standard cases, these charges are conserved, $dQ^a/dt=0$, and they have a well defined commutation relations given by

\begin{equation}
[Q^a,Q^b]=C^{abc}Q^c,
\end{equation}
where $C^{abc}$ are the structure constants of the Lie algebra. We can define an unitary operator with the charge being the generator of the group transformations

\begin{equation}
U=e^{iQ^a\alpha^a}.
\end{equation}
If the vacuum is non-degenerate, then the previously defined charge annihilates the vacuum, namely, $U\vert0>=\vert0>$, or equivalently

\begin{equation}
Q^a\vert0>=0.
\end{equation}    
When the vacuum is degenerate, then these previous conditions are not satisfied and in general

\begin{equation}
U\neq e^{iQ^a\alpha^a}, \;\;\;\;\;Q^a\vert0>\neq0. 
\end{equation}
If the operator $\phi(x)$ is not a singlet, then its commutation with the charge $Q^a$ is non-zero and given by

\begin{equation}   \label{phihi}
[Q^a,\phi'(x)]\backsim\phi(x).
\end{equation}    
Later we will see that the right-hand side will come out to be the order parameter of the system. When the symmetry is spontaneously broken, then the vacuum expectation value of the order parameter does not vanish for an arbitrarily selected vacuum. Then the result (\ref{phihi}) gives

\begin{equation}   \label{phihi2}
<0\vert Q^a\phi'(x)-\phi'(x)Q^a\vert0>\neq0.
\end{equation}
If we introduce the definition of charge given in eq. (\ref{eq:Chargedef}), then we get

\begin{equation}
\sum_n\int d^3y\left[<0\vert j^a_0(y)\vert n><n\vert\phi'(x)\vert0>-<0\vert\phi'(x)\vert n><n\vert j^a_0(y)\vert0>\right]_{x^0=y^0}\neq0,
\end{equation}
where the equal-time condition has been imposed. Under spacetime translational invariance, this previous expression finally becomes

\begin{equation}
(2\pi)^3\sum_n\delta^{(3)}(\vec{p}_n)\left[<0\vert j^a_0(0)\vert n><n\vert\phi'(x)\vert0>e^{iM_ny_0}-<0\vert\phi'(x)\vert n><n\vert j^a_0(0)\vert0>e^{-iM_ny^0}\right]_{x^0=y^0},
\end{equation}
which must be different from zero. Note that $p_n^0=M_n$ and the spatial integrals were evaluated, obtaining in this way the delta-Diracs. The current conservation guarantees that the previous expression is independent of $y^0$. Then it is trivial to observe that $M_n=0$ and this proves the Nambu-Goldstone theorem because the particles are gapless and the momentum and frequency go to zero simultaneously. From gapless particles we would expect a linear dispersion relation. The previous demonstration also suggests that every broken symmetry is related to the existence of one Nambu-Goldstone boson. The proof elaborated in the previous way can be found in \cite{Field2}. This method cannot say directly the relation between the number of broken symmetries and the number of Nambu-Goldstone bosons in the most general situations. Later we will see how we can extract more information from the basic expression (\ref{phihi}). For this purpose, we can rewrite the results (\ref{phihi}) and (\ref{phihi2}) by using bi-linear objects to be justified later. Then the Nambu-Goldstone theorem at the quantum level can be formulated suggesting that given a field $\phi_{a,b}(x)$ which is not a singlet under the action of the generator of a group, then its vacuum expectation value satisfies the condition

\begin{equation}   \label{eq:eq1}
<0_{SV}\vert \phi_{a,b}(x)\vert0_{SV}>\neq0.
\end{equation}  
The sub-index $SV$ means single vacuum. Note that here we take the field $\phi_{ab}(x)$ as a second rank tensor instead of a vector or scalar. We will consider the general situation where $\phi_{a,b}(\vec{x})=\phi_{a,b}^{m,l}\epsilon_m\otimes\epsilon_l$. Since the field is not a singlet under the action of the broken generators, then it satisfies the condition

\begin{equation}   \label{eq:eq2}
[Q_{m,k}(y),\phi_{a,b}(x)]\backsim\phi'_{a,b}(x).
\end{equation}   
Note that here we also take the conserved charges as a second order rank tensors. By combining the two conditions, namely, eqns. (\ref{eq:eq1}) and (\ref{eq:eq2}), we obtain the result

\begin{equation}   \label{eq:eq3}
<0_{SV}\vert [Q_{m,k}(y),\phi_{a,b}(x)]\vert0_{SV}>\neq0.
\end{equation}
Here $Q_{m, k}$ corresponds to conserved charges. In general, we can take $Q_{m,l}=Q_{m,l}^{p,k}\epsilon_p\otimes \epsilon_k$. Eq. (\ref{eq:eq3}), is an equation with two pairs of indices living in different spaces respectively. The meaning of the indices will become clear later.    

\section{The Quantum Yang-Baxter conditions}   \label{eq:Sec2}

The QYBE are defined by the relation

\begin{equation}   \label{amazing}
R_{(1,2)}R_{(1,3)}R_{(2,3)}=R_{(2,3)}R_{(1,3)}R_{(1,2)},
\end{equation}
in operator notation. The matrices will be bi-linear objects acting on a space $M\otimes V$. In coordinate notation, we can define the matrices as follows \cite{The book}

\begin{equation}   \label{99}
R:M\otimes V=R^{a,b}_{c,d}m_a\otimes m_b.
\end{equation}
Analogous conclusions appear for the other matrices. There might be cases where the spaces $M$ and $V$ are the same. Here $B=\{m_1, m_2, ..., m_n\}$ is a basis of $M$. The space $V$ will have an analogous basis. In coordinate notation, the QYBE defined in eq. (\ref{amazing}), become \cite{The book}

\begin{equation}   \label{This is the form}
R^{s_2, s_3}_{j,  k}R^{s_1, c}_{i, s_3}R^{a,b}_{s_1,s_2}=R^{r_1,r_2}_{i,j}R^{a,r_3}_{r_1,k}R^{b,c}_{r_2,r_3}.
\end{equation}
In this paper, we will use this coordinate form for the QYBE.   

\subsection{The relations between the order parameter and the conserved charges}   \label{NS}

\begin{figure}
	\centering
	\includegraphics[width=10cm, height=8cm]{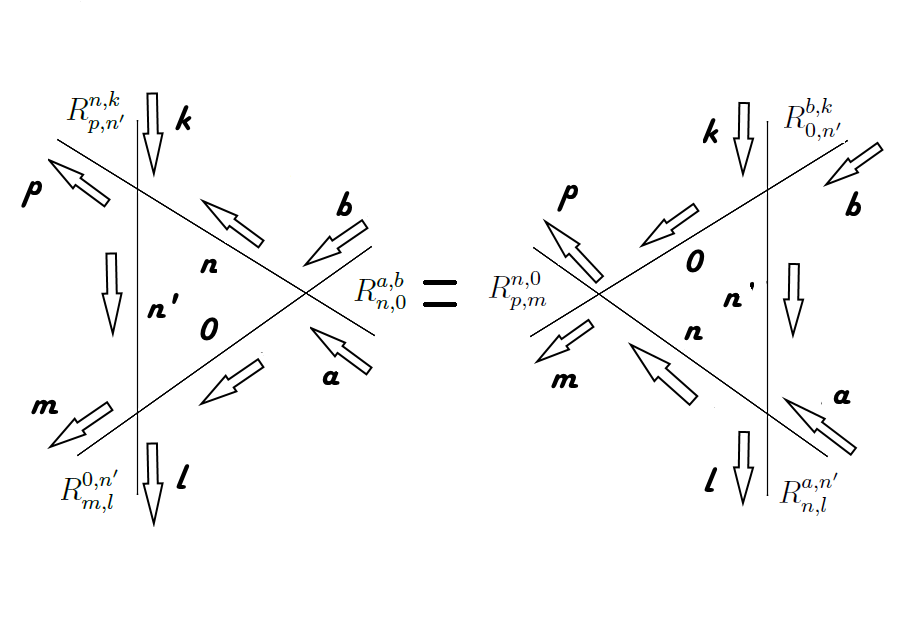}
	\caption{The QYBE defined by eq. (\ref{eq:the last option2}).}
	\label{The titan3}
\end{figure}

The QYBE defined in eq. (\ref{This is the form}), admits the following form

\begin{equation}   \label{eq:the last option2}
R^{0,n'}_{m,l}R^{n,k}_{p,n'}R^{a,b}_{n,0}=R^{n,0}_{p,m}R^{a,n'}_{n,l}R^{b,k}_{0,n'},
\end{equation}
which corresponds to an equation for the product of three matrices defined by

\begin{eqnarray}   \label{another form3}
R^{0,n'}_{m,l}=<0_{DV}\vert Q_{m,l}(y)\vert n'>,\;\;\;R^{n,k}_{p,n'}=<n'\vert Q_{k,p}(z)\vert n>,\;\;\;\nonumber\\
R^{a,b}_{n,0}=<n\vert \phi_{a,b}(x)\vert0_{DV}>,\;\;\;R^{n,0}_{p,m}=<n\vert Q_{p, m}(z)\vert0_{DV}>,\;\;\;\nonumber\\
R^{a,n'}_{n,l}=<n'\vert Q_{l, a}(y) \vert n>,\;\;\;\;R^{b,k}_{0,n'}=<0_{DV}\vert \phi_{b, k}(x)\vert n'>.\;\;\;\;\;
\end{eqnarray} 
Note that the index notation is consistent. This can be observed from the figure \ref{The titan3}. In the figure, each line in the triangles, represent one space. Then for example, the index group $b$, $0$ and $m$ lives in the same space and the group of index represented by $k$, $n'$ and $l$ lives in another space. This statement appears consistently in the matrix $R^{0,n'}_{m,l}$, where $0$ and $m$, appearing vertically one over the other, implies that they are connected to the same space. The same logic applies to the pair $n'$ and $l$ in the same matrix. The same consistency appears if we repeat the same reasoning over the matrix $R^{b,k}_{0,n'}$ in eq. (\ref{another form3}). The same analysis can be repeated over all the other matrices. It is important to understand that we can only contract pairs of index living in the same space. This can be seen clearly from the QYBE in eq. (\ref{eq:the last option2}). The contracted group of index, represent either, gapless intermediate particles or the degenerate vacuum and this means that the QYBE demand us to sum over all the possible vacuums. On the other hand, the free indices like $b$, $p$ and $l$, are the labels of the Nambu-Goldstone field and the pair of broken generators. Since $b$ is connected to the same space of the index $0$, then this means that the Nambu-Goldstone field will always appear operating over the vacuum in the different expressions. Repeating the same reasoning, the index $p$ is the label of the broken generator related to the modes $n$ and the index $l$ is the label for the broken generator related to the modes $n'$. Finally, the indices $a$, $k$ and $m$ are auxiliary. They are just telling us where should be located the modes $n$, $n'$ as well as the degenerate vacuum when we expand the different expressions in the commutator representing the spontaneous symmetry breaking condition. This point will be clearer in a moment. Look for example that the index $a$ is always tied to the mode $n$. Then whenever it appears is indicating us that the mode $n$ appears as a bra or as a ket for the corresponding operator. From eq. (\ref{eq:the last option2}), the following equality is satisfied

\begin{eqnarray}   \label{another form2}   
\sum_{0, n, n'}<0_{DV}\vert Q_{m,l}(0)\vert n'><n'\vert Q_{k,p}(0)\vert n><n\vert \phi_{a,b}(x)\vert0_{DV}>=\sum_{0, n, n'}<0_{DV}\vert \phi_{b, k}(x)\vert n'>\times\nonumber\\
<n'\vert Q_{l, a}(0) \vert n><n\vert Q_{p, m}(0)\vert0_{DV}>.
\end{eqnarray} 
The sub-index $DV$ means {\it degenerate vacuum}. We have introduced pairs of complete set of intermediate states defined by $\hat{I}=\sum_n\vert n><n\vert=\sum_{n'}\vert n'><n'\vert$. In addition, we are summing over the multiplicity of vacuums $\vert 0_{DV}>$. We will justify this operation at the end of this paper. The sum over the vacuums is represented by the contraction between the indices $0$ in eq. (\ref{eq:the last option2}). Note that the vacuum will then appear as internal lines in the graphical version of the QYBE as it is shown in Fig. \ref{The titan3}. Here we will further explain the meaning of the free indices. The free indices corresponding to the Nambu-Goldstone field $\phi_{ab}$ and broken generators, are indices living in the corresponding spaces of the Goldstone bosons or the degenerate vacuum. For example, in the matrix $R^{n,k}_{p,n'}$, the index $p$ lives in the space of the family of Goldstone bosons represented by $n$, meanwhile, the index $k$ lives in the space of the family of Goldstone bosons represented by $n'$. This matrix then represents the interaction of pairs of Goldstone bosons ($n$ and $n'$) and we can perceive the pair of indices $k$ and $p$ as the labels representing such interaction in the vertex of the Yang-Baxter diagram. Consider now the Nambu-Goldstone field $\phi_{a,b}$ in eq. (\ref{another form2}). In this case, the pair of indices $a,b$ represent the interaction between the Goldstone bosons $n$ with the vacuum $0_{DV}$. Then we can think on $\phi_{a,b}$ as the Nambu-Goldstone field touching the legs corresponding to $n$ and $0_{DV}$ in the diagrams. Analogous conclusions apply to the broken generators. In order to agree with the ordinary commutators, we will take as the fundamental labels for the broken generators those indices living in the spaces corresponding to the different families of the Goldstone bosons. If one broken generator has two indices living in different families of Goldstone bosons $n$ and $n'$, then we will take as the fundamental label the index living in the space belonging to the family of Goldstone bosons which is not a common vertex with respect to another broken generator. Then for example, in eq. (\ref{another form2}), we will consider the indices $l$ and $p$ as the real labels for the broken generators. Then $Q_{m, l}=Q_l$ and $Q_{l, a}=Q_l$ will represent the same broken generators in eq. (\ref{another form2}). Analogously, $Q_{p, m}=Q_p$ and $Q_{k,p}=Q_p$ will also correspond to a single broken generator. For the case of the Nambu-Goldstone field, the situation is different, the fundamental index is the one living in the vacuum space. Then $\phi_{a,b}=\phi_{b, k}=\phi_b$ and in such a case the fundamental index will be $b$, which lives in the space $0_{DV}$. Then in this way, eq. (\ref{another form2}) could be expressed in a simplified notation as

\begin{eqnarray}   \label{king form2} 
\sum_{0}<0_{DV}\vert Q_l(0)Q_p(0)\phi_b(x)\vert0_{DV}>=\sum_{0}<0_{DV}\vert\phi_b(x) Q_l(0)Q_p(0)\vert0_{DV}>.
\end{eqnarray}

\subsubsection{Exchange of the intermediate particles $n\to n'$}

If we exchange the spaces containing the intermediate states $n\to n'$, then eq. (\ref{eq:the last option2}) is transformed into

\begin{equation}   \label{eq:the last optionmiau}
R^{0,n}_{m,p}R^{n',a}_{l,n}R^{k,b}_{n',0}=R^{n',0}_{l,m}R^{k,n}_{n',p}R^{b,a}_{0,n},
\end{equation}
or in operator notation, eq. (\ref{amazing}) becomes

\begin{equation}   \label{amazing2}
R_{(3,2)}R_{(3,1)}R_{(2,1)}=R_{(2,1)}R_{(3,1)}R_{(3,2)}.
\end{equation}
This equation is the twist map of eq. (\ref{amazing}). Under the exchange of spaces, the pair of indices $\{(0,m),(n,p),(n',l),(a,n),(k,n'),(b,0)\}$ remain together if we compare this case with the one analyzed in eq. (\ref{eq:the last option2}). This is equivalent to say that we only move the lines with respect to each other in the QYB diagrams as can be observed from Fig. \ref{Theking2}. Eq. (\ref{eq:the last optionmiau}) represents the following equality

\begin{eqnarray}   \label{eq:the last optionmiaunoagain}
\sum_{0, n, n'}<0_{DV}\vert Q_{m,p}(0)\vert n><n\vert Q_{a,l}(0)\vert n'><n'\vert \phi_{k,b}(x)\vert0_{DV}>=\sum_{0, n, n'}<0_{DV}\vert \phi_{b, a}(x)\vert n>\times\nonumber\\
<n\vert  Q_{p, k}(0)\vert n'><n'\vert Q_{l, m}(0)\vert0_{DV}>,
\end{eqnarray}
if we use an analogous notation as in eq. (\ref{another form3}). If $n=n'$, then eqns. (\ref{amazing}) and (\ref{amazing2}) are the same, except for the fact that they are the time reversal version of each other and this will be reflected in the corresponding phases. If $n\neq n'$, then the mentioned equations are different. In simplified notation, the previous result can be expressed as

\begin{eqnarray}   \label{king form2la} 
\sum_{0}<0_{DV}\vert Q_p(0)Q_l(0)\phi_b(x)\vert0_{DV}>=\sum_{0}<0_{DV}\vert \phi_b(x) Q_p(0)Q_l(0)\vert0_{DV}>.
\end{eqnarray}   

\section{The spontaneous symmetry breaking condition and the Yang-Baxter relations}   \label{extension}

In general, if we have at least two or more different broken generators and if they satisfy a Lie algebra, then we can re-express the spontaneous symmetry breaking condition defined in eq. (\ref{eq:eq3}) as follows

\begin{equation}   \label{impo}
<0_{SV}\vert \left[\phi_{b,a}(x), [Q_{p,k}(y),Q_{l,m}(z)]\right]\vert0_{SV}>\neq0.
\end{equation}
Under the assumption of invariance under space-time translations defined through the equality $Q_{k,p}(y)=e^{-ipy}Q_{k, p}(0)e^{ipy}$, the previous commutator gives an expression of the form

\begin{eqnarray}   \label{nojoda}
\sum_{n, n'}<0_{SV}\vert\phi_{b}(x)\vert n><n\vert\ Q_{p}(0)\vert n'><n'\vert Q_{l}(0)\vert0_{SV}>e^{-i(p_n-p_{n'})y}e^{-i\tilde{p}_{n'}z}\nonumber\\
-<0_{SV}\vert\phi_{b}(x)\vert n'><n'\vert\ Q_{l}(0)\vert n><n\vert Q_{p}(0)\vert0_{SV}>e^{-i(\tilde{p}_{n'}-\tilde{p}_n)z}e^{-ip_{n}y}\nonumber\\
-<0_{SV}\vert\ Q_{p}(0)\vert n><n\vert Q_{l}(0)\vert n'><n'\vert\phi_{b}(x)\vert 0_{SV}>e^{-i(\tilde{p}_{n}-\tilde{p}_{n'})z}e^{ip_{n}y}\nonumber\\
+<0_{SV}\vert Q_{l}(0)\vert n'><n'\vert\ Q_{p}(0)\vert n><n\vert\phi_{b}(x)\vert0_{SV}>e^{-i(p_{n'}-p_{n})y}e^{i\tilde{p}_{n'}z}\neq 0.
\end{eqnarray}
Here $Q_{p,k}=Q_{p,m}=Q_{m,p}=Q_{k,p}=Q_p$; $Q_{l,m}=Q_{l,a}=Q_{a,l}=Q_{m,l}=Q_l$, and $\phi_{b, a}=\phi_{b, k}=\phi_{k, b}=\phi_{a, b}=\phi_b$ in order to write the previous expression in a simplified notation. Here we define $p_n$ as the 4-momentum for the mode associated to one of the conserved charges. Meanwhile, $\tilde{p}_n$ is the 4-momentum associated to the other operator in the same algebra. The equality $p_n=p_{n'}$ as well as $\tilde{p}_{n}=\tilde{p}_{n'}$ comes from spatial integrations after expressing the broken generators as density integrations $Q_{b}(y)=\int d^3y j_{b}(y)$. These conditions are then independent of the QYBE and they appear in the thermodynamic limit \cite{Tomas, Tomas4}. In what follows, we will develop the expressions for two cases. Note that $e^{ipy}$ is a unitary operator related to spacetime translations and not an ordinary number. Note in addition that $y=(y^0, {\bf y})$ corresponds to the spacetime index.

\subsubsection{Pairs of Goldstone bosons representing the same degree of freedom}

\begin{figure}
	\centering
		\includegraphics[width=10cm, height=8cm]{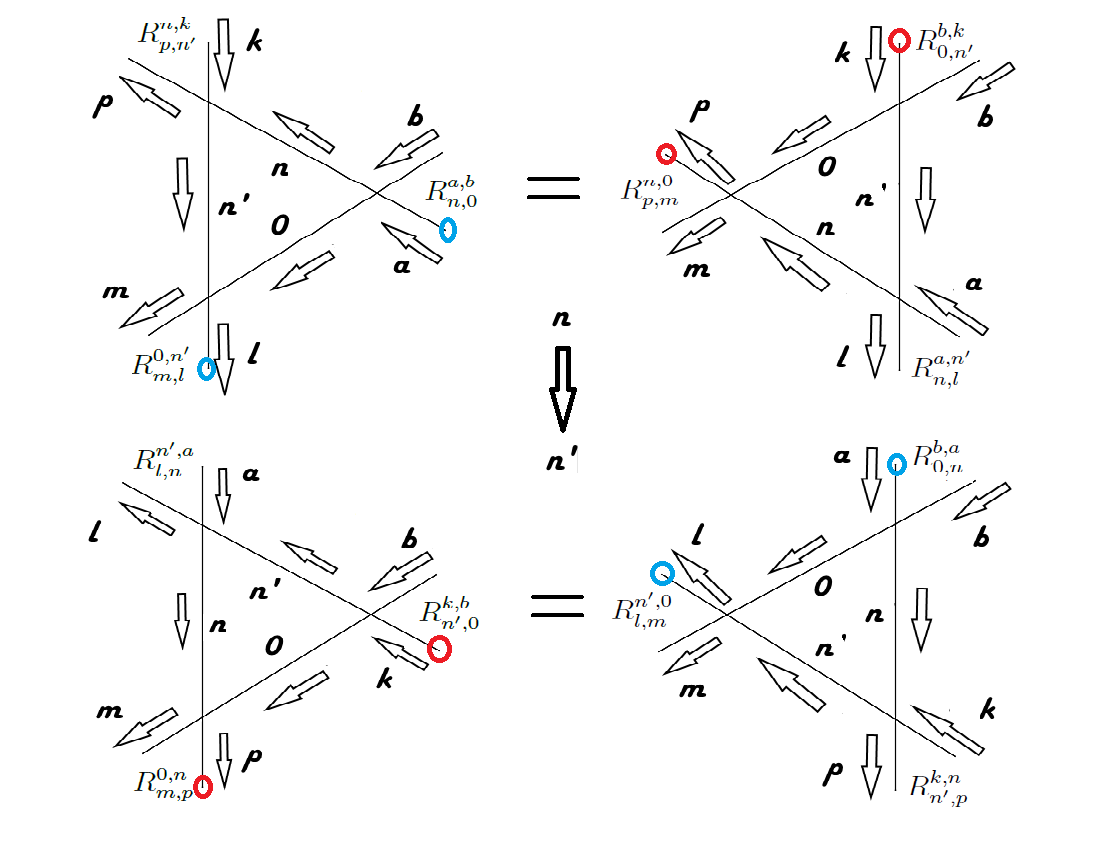}
	\caption{The effect of exchanging $n\to n'$. The upper relation, corresponds to the QYBE defined in eq. (\ref{eq:the last option2}). The lower relation corresponds to eq. (\ref{eq:the last optionmiau}). The figures marked with the same color are twist map (mirror image) of each other. The arrows represent the flow of time for each phase.}
	\label{Theking2}
\end{figure}

\begin{figure}
	\centering
		\includegraphics[width=10cm, height=8cm]{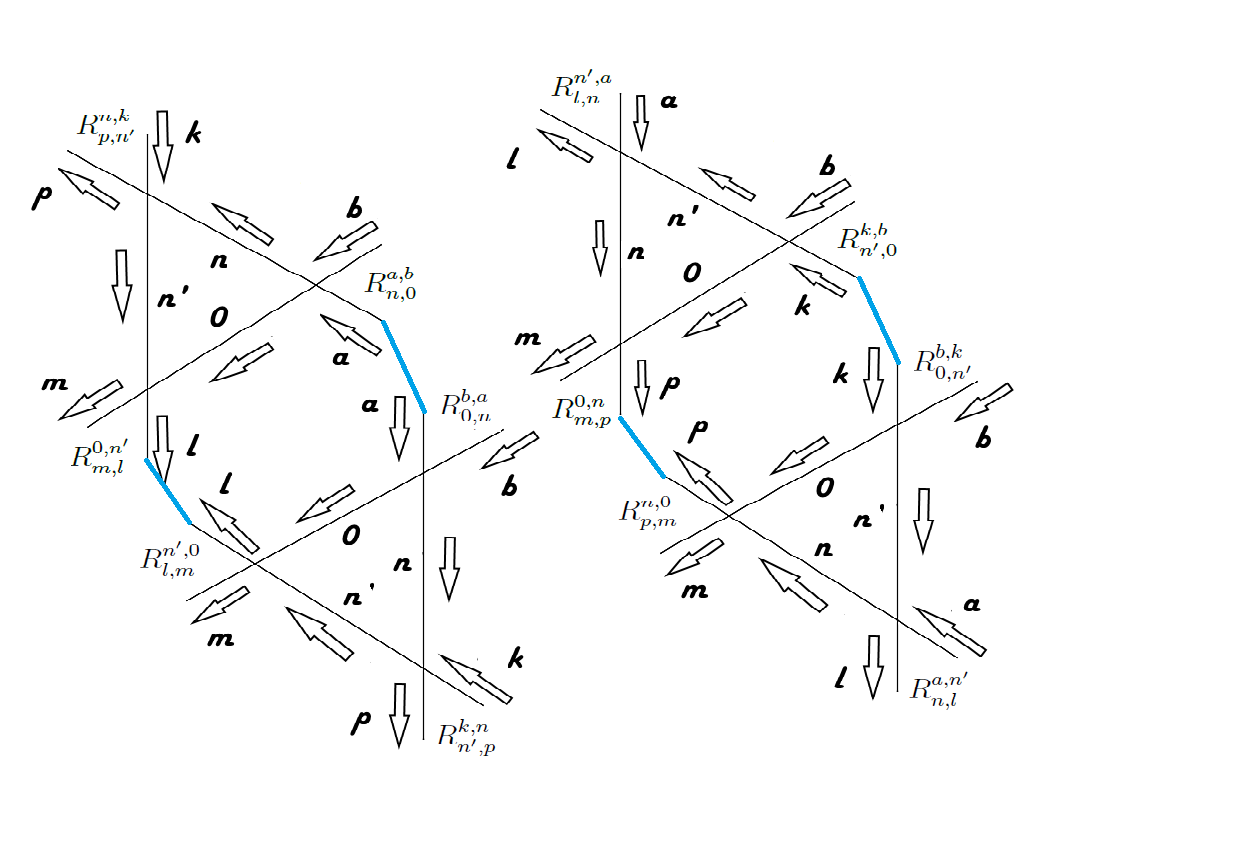}
	\caption{The effect of exchanging $n$ by $n'$. The figures connected by blue lines, are mirror image of each other.}
	\label{Theking3}
\end{figure}

In this case, besides the conditions $p_n=p_{n'}$ and $\tilde{p}_n=\tilde{p}_{n'}$, the condition $n=n'$ must be satisfied. Then eqns. (\ref{king form2}) and (\ref{king form2la}) are the same. After summing over the degenerate vacuum and factorizing terms with common coordinates, then eq. (\ref{nojoda}) can be reduced to
\begin{eqnarray}   \label{nojoda3}
\sum_{0, n, n'}<0_{DV}\vert\phi_{b}(x)\vert n><n\vert\ Q_{p}(0)\vert n'><n'\vert Q_{l}(0)\vert0_{DV}>2e^{-i\tilde{E}_{n'}z_0}cos\left(\tilde{{\bf p}}_n\cdot{\bf z}\right)\nonumber\\
-<0_{DV}\vert\ Q_{p}(0)\vert n><n\vert Q_{l}(0)\vert n'><n'\vert\phi_{b}(x)\vert 0_{DV}>2e^{iE_{n}y_0}cos\left({\bf p}_n\cdot{\bf y}\right)=0.
\end{eqnarray}
where we have used the result $p_ny=E_ny_0-\vec{p_n}\cdot\vec{y}$, as well as the fact that both of the QYBE defined in eqns. (\ref{eq:the last option2}) and (\ref{eq:the last optionmiau}) are the same under the condition $n=n'$. Note that at the moment of grouping common factors in eq. (\ref{nojoda3}), we have taken into account that eq. (\ref{king form2la}) is the twist map of eq. (\ref{king form2}) and as a consequence both equations are time reversed with respect to each other. This can be observed from Figs. \ref{Theking2} and \ref{Theking3}, where the arrows in the internal lines represent the flow of the corresponding phase. Later we will see how the phase convention can be arranged naturally just following the QYBE in coordinate notation. Here as a consequence of the condition $n=n'$, the momentum of the intermediate particles approaches to zero quadratically at the lowest order in the expansion. From the time-independence condition of eq. (\ref{nojoda3}), the energy (frequency) approaches to zero linearly. Note that here $p_n=\tilde{p}_n$, as well as $E_n=\tilde {E}_{n'}$ due to the condition $n=n'$ applied in eq. (\ref{nojoda}). By applying the QYBE as they were defined in eq. (\ref{king form2la}) in simplified notation, the previous result can be expressed as

\begin{eqnarray}   \label{nojoda5}
2\sum_{0, n, n'}<0_{DV}\vert\phi_{b}(x)\vert n><n\vert\ Q_{p}(0)\vert n'><n'\vert Q_{l}(0)\vert0_{DV}>(e^{-i\tilde{E}_{n'}z_0}cos\left(\tilde{{\bf p}}_n\cdot{\bf z}\right)\times\nonumber\\
-e^{iE_{n}z_0}cos\left({\bf p}_n\cdot{\bf y}\right))=0,\;\;\;\;\;\;\;
\end{eqnarray}
under the equal time-commutation condition $z_0=y_0$. In the limit ${\bf y}\to {\bf z}$, we have

\begin{eqnarray}   \label{nojodadelajoda5}
4i\sum_{0, n, n'}<0_{DV}\vert\phi_{b}(x)\vert n><n\vert\ Q_{p}(0)\vert n'><n'\vert Q_{l}(0)\vert0_{DV}>sin(E_n z_0)cos\left({\bf p}_n\cdot{\bf z}\right)_{{\bf y}\to {\bf z}}=0.
\end{eqnarray}
Here we can see that $E_n\to0$ linearly and $p_n\to0$ quadratically. The previous result is consistent with the notion of trace of commutators. We can then conclude that the interaction of two identical Nambu-Goldstone bosons has a direct correspondence with the QYBE if we sum over the degenerate vacuum.     

\subsubsection{Pairs of Goldstone bosons independent}

When the Goldstone bosons are independent, we can still write the general result (\ref{nojoda}) since it represents the standard spontaneous symmetry breaking condition. In this case however, since $n\neq n'$, then we cannot claim that eqns. (\ref{eq:the last option2}) and (\ref{eq:the last optionmiau}) are the same. Then there are some pair of terms in the general expression (\ref{nojoda}) that we cannot factorize. This at the end will turn out to be the mathematical reason for getting a linear dispersion relation corresponding to the Nambu-Goldstone bosons. By following similar arguments as in the previous case, we obtain the result
 
\begin{eqnarray}   \label{nojodadelajodajoda22}
\sum_{0, n, n'}<0_{DV}\vert\phi_{b}(x)\vert n><n\vert\ Q_{p}(0)\vert n'><n'\vert Q_{l}(0)\vert0_{DV}>(e^{-i(p_n-p_{n'})y}e^{-i\tilde{p}_{n'}z}\nonumber\\
-e^{-i(\tilde{p}_{n}-\tilde{p}_{n'})z}e^{ip_{n}y})-<0_{DV}\vert\phi_{b}(x)\vert n'><n'\vert\ Q_{l}(0)\vert n>\times\nonumber\\
<n\vert Q_{p}(0)\vert0_{DV}>(e^{-i(\tilde{p}_{n'}-\tilde{p}_n)z}e^{-ip_{n}y}-e^{-i(p_{n'}-p_{n})y}e^{i\tilde{p}_{n'}z})=0.
\end{eqnarray}
Note that here we have used the QYBE in simplified notation defined in eqns. (\ref{king form2}) and (\ref{king form2la}) in order to factorize some terms in eq. (\ref{nojoda}). In this case these two pair of equations, namely, (\ref{king form2}) and (\ref{king form2la}) are completely independent. Then we cannot factorize all the terms in eq. (\ref{nojodadelajodajoda22}) as we did for the previous case. For this reason, in the neighborhood when the momentum and frequency go simultaneously to zero, the dispersion relation is linear $E_n\backsim {\bf p}_n$.    
   											
\section{The counting of Nambu-Goldstone bosons based on the Yang-Baxter relations}   \label{The counting}

Since the $R$-matrices represent interactions between $n$, $n'$ and $0_{DV}$, then their rank will be related to the number of Goldstone bosons. Here we have square matrices, then we will have as many degenerate vacuums ($0_{DV}$) as Goldstone bosons the system under study has. This is consistent with the fact that the action of a broken generator over one vacuum is just to rotate it to a different one. This is demonstrated in \cite{Tomas4} by considering coherent states for the vacuum. In our analysis if $n=n'$, then we have $Rank (R)=n+n'=2n=N_{BG}$ and then $N_{NG}=(1/2)N_{BG}$. If $n\neq n'$, then $N_{BG}=N_{NG}$. The general relation between the number of Nambu-Goldstone bosons and the number of broken generators is defined by $N_{NG}=\frac{1}{2}Rank(R^{i,j}_{k,l})_{n=n'}+Rank(R^{i,j}_{k,l})_{n\neq n'}$. Here the first matrix $R_{n=n'}$ corresponds to the matrix formed for the case where pairs of Goldstone bosons represent the same degree of freedom $n=n'$. The second matrix $R_{n\neq n'}$, corresponds to the case where the Goldstone bosons are completely independent from each other. Here we can classify the Nambu-Goldstone bosons as ${\bf type A}:n_A=Rank(R^{i,j}_{k,l})_{n\neq n'}$ and ${\bf Type B}:n_B=\frac{1}{2}Rank(R^{i,j}_{k,l})_{n=n'}$. Then $N_{NG}=n_A+n_B$ and $N_{BG}=Rank(R^{i,j}_{k,l})_{n=n'}+Rank(R^{i,j}_{k,l})_{n\neq n'}=2n_B+n_A$ as it should be. The final formula relating $N_{BG}$ with $N_{NG}$ is defined in terms of the $R$-matrices as follows

\begin{equation}   \label{counting}
N_{NG}=N_{BG}-\frac{1}{2}Rank(R^{i,j}_{k,l})_{n=n'}.
\end{equation}
Note the analogy with the situation described in \cite{Murayama}. However, in this paper we remark that the fundamental Mathematical objects are the $R$-matrices and all the dynamic is governed by the QYBE. Then although the approach developed here is different to the one done in \cite{Murayama}, in the sense that it focuses on the dynamical constraints appearing from the QYBE, analogous results with respect to previous approaches still appear, showing then consistency. Then at least in the sense of the counting of Nambu-Goldstone bosons, the role of the $\rho$ matrix in \cite{Murayama}, is assumed by the rank of the $R$-matrices here. Alternatively, in the present formalism, the number of Nambu-Goldstone bosons can be found naturally from the number of independent triangles (see the Figure \ref{Theking2}) appearing from the Yang-Baxter expressions.            

\section{Sum of vacuums and theorems connected with the spontaneous symmetry breaking phenomena}   \label{Hiddenla?}

\begin{figure}
	\centering
		\includegraphics[width=16cm, height=8cm]{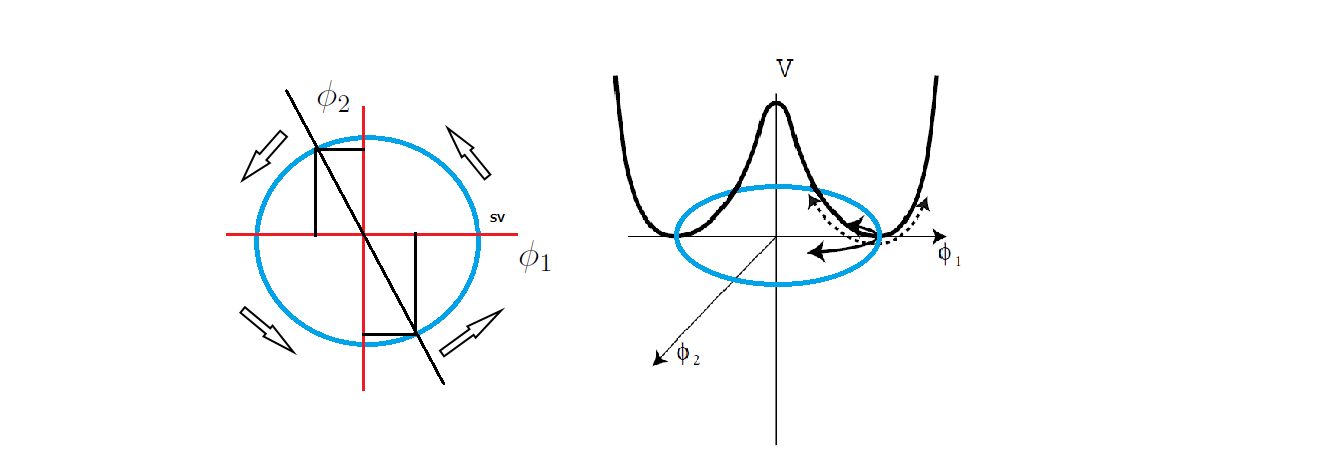}
	\caption{The degenerate vacuum corresponding to the Nambu-potential. The Black line crossing the circumference suggests that for each positive value of the order parameter on a given vacuum (expectation value), there is a corresponding negative value with the same magnitude but defined in a different vacuum. Then summing over the expectation values of the of the order parameters, obtained along the degenerate vacuum gives a trivial (vanishing) result. The arrows illustrate the direction of the sum. The right-hand side of the figure is the potential term defined in eq. (\ref{pot}) and the left-hand side is the same potential observed over the $\phi_1-\phi_2$ plane. The right-hand side is partially taken from \cite{Tomas3} and briefly modified.}
	\label{Iwillbeback}
\end{figure}

In the previous sections we have demonstrated that the Yang-Baxter relations contain the Nambu-Goldstone theorem and they also guarantee the appropriate counting of the Nambu-Goldstone bosons. The correct dispersion relations are also obtained naturally. In this section we will analyze this issue in more detail and we will formulate a fundamental theorem related to the dispersion relations and the counting of the Nambu-Goldstone bosons. We will also justify why the Yang-Baxter relations can be really considered as the appropriate formulation of the spontaneous symmetry breaking phenomena and we will explain why it is possible to sum over the degenerate vacuum even if each vacuum expands in principle a different Hilbert space. We start with the Lagrangian 

\begin{equation}   \label{Thislagrangian}
\pounds=\partial^\mu\phi^*\partial_\mu\phi-V(\phi^*\phi),
\end{equation}
where the potential is defined as 

\begin{equation}   \label{pot}
V(\phi\phi^*)=-\mu^2\phi\phi^*+\lambda(\phi\phi^*)^2, \;\;\;\;\;\lambda>0.
\end{equation}    
The Lagrangian (\ref{Thislagrangian}) is invariant under the global $U(1)$ transformation defined as $\phi\to e^{i\alpha}\phi$. We can extract from the full field $\phi$ the Nambu-Goldstone degree of freedom plus the order parameter field if we define

\begin{equation}
\phi=\frac{1}{\sqrt{2}}(\phi_1+i\phi_2).
\end{equation}    
We can identify $\phi_2$ as the Nambu-Goldstone mode with zero vacuum expectation value, namely $<0_{SV}\vert\phi_2\vert0_{SV}>=0$. The order parameter then breaks the symmetry spontaneously as

\begin{equation}   \label{Thisexpression}
<0_{SV}\vert\phi_1\vert0_{SV}>=\pm\mu/\sqrt{\lambda}.
\end{equation}  
\subsection{The justification behind the sum over the vacuums}

Note that in the previous example for every positive value of the order parameter, there exists a negative (averaged) value with the same magnitude. The sign taken by the order parameter as well as its magnitude, defines the selected vacuum. It is evident then that a sum over all possible vacuums is in agreement with the expression (\ref{Thisexpression})

\begin{equation}   \label{Thisonela}
\sum_{0}<0_{DV}\vert\phi_1\vert0_{DV}>=\bar{\phi}_1-\bar{\phi}_1+\bar{\phi}_2-\bar{\phi}_2+...+\bar{\phi}_n-\bar{\phi}_n=\sum_{n=1}^{N}\bar{\phi}_i=0.
\end{equation}
The spontaneous symmetry breaking condition suggests that the vacuum expectation value of the commutator between the Nambu-Goldstone field and the broken generator is proportional to the order parameter   

\begin{equation}   \label{Thisonela2}
<0_{SV}\vert[Q, \phi_2]\vert0_{SV}>=i<0_{SV}\vert\phi_1\vert0_{SV}>.
\end{equation}
From eq. (\ref{Thisonela}) and (\ref{Thisonela2}), we can conclude that

\begin{equation}   \label{here la}
\sum_{0}<0_{DV}\vert[Q, \phi_2]\vert0_{DV}>=0.
\end{equation}
The figure \ref{Iwillbeback} illustrates the essence of this result. The vertical axis on the left-hand side represents the value taken by the field $\phi_2$ and the horizontal axis represents the value taken by the field $\phi_1$. Each point in the circumference represents a vacuum state condition. If we select $\phi_2$ to be the Nambu-Goldstone field, then its vacuum expectation value vanishes and the field $\phi_1$ has to be massive. This result corresponds to the point marked in the figure \ref{Iwillbeback} as $SV$. Note that the result (\ref{here la}), is equivalent to the expansion (\ref{impo}) after summing over the degenerate vacuum. Eq. (\ref{here la}) is also equivalent to the QYBE as can be seen from the definitions given in eq. (\ref{eq:the last option2}), which is equivalent to eq. (\ref{king form2}) and from eq. (\ref{eq:the last optionmiau}), which is equivalent to eq. (\ref{king form2la}). Then the QYBE suggest that the sum of the expectation values of the order parameter over all the possible vacuums is equal to zero. An equivalent and more evident interpretation is that once we sum over the degenerate vacuum, it does not matter what is the order of operation over the vacuum states. This means that we can operate either, first with the Nambu-Goldstone field and after with the broken generator or first with the broken generator and after with the Nambu-Goldstone field, obtaining then identical results. We can conclude that the QYBE are naturally satisfied by the systems where the symmetry is spontaneously broken before any selection of the vacuum is done. Once the vacuum is selected due to small perturbations, in principle the QYBE are not satisfied. However, the results obtained before suggest that the Nambu-Goldstone bosons still keep some memory on the fact that the real vacuum should be a superposition of all the possible available vacuums. The dispersion relations of the Nambu-Goldstone bosons are then the evidence of this result. This statement makes sense because in fact, the Nambu-Goldstone bosons move along different vacuums and only an external fluctuation is able to select some state. The Nambu-Goldstone field itself, does not mind what kind of vacuum is selected. In other words, what defines the correct dispersion relation of the Nambu-Goldstone bosons is the space through which they can move and this is contained inside the QYBE. It might be surprising to see the fact that we have to sum over the degenerate vacuum since each vacuum represents a different Hilbert space in the thermodynamic limit (infinitie volume limit) \cite{Tomas, Tomas4}. This is not the case however in the finite volume limit. 

\subsubsection{Further justifications of the sum over the degenerate vacuum}

\begin{figure}
	\centering
		\includegraphics[width=14cm, height=8cm]{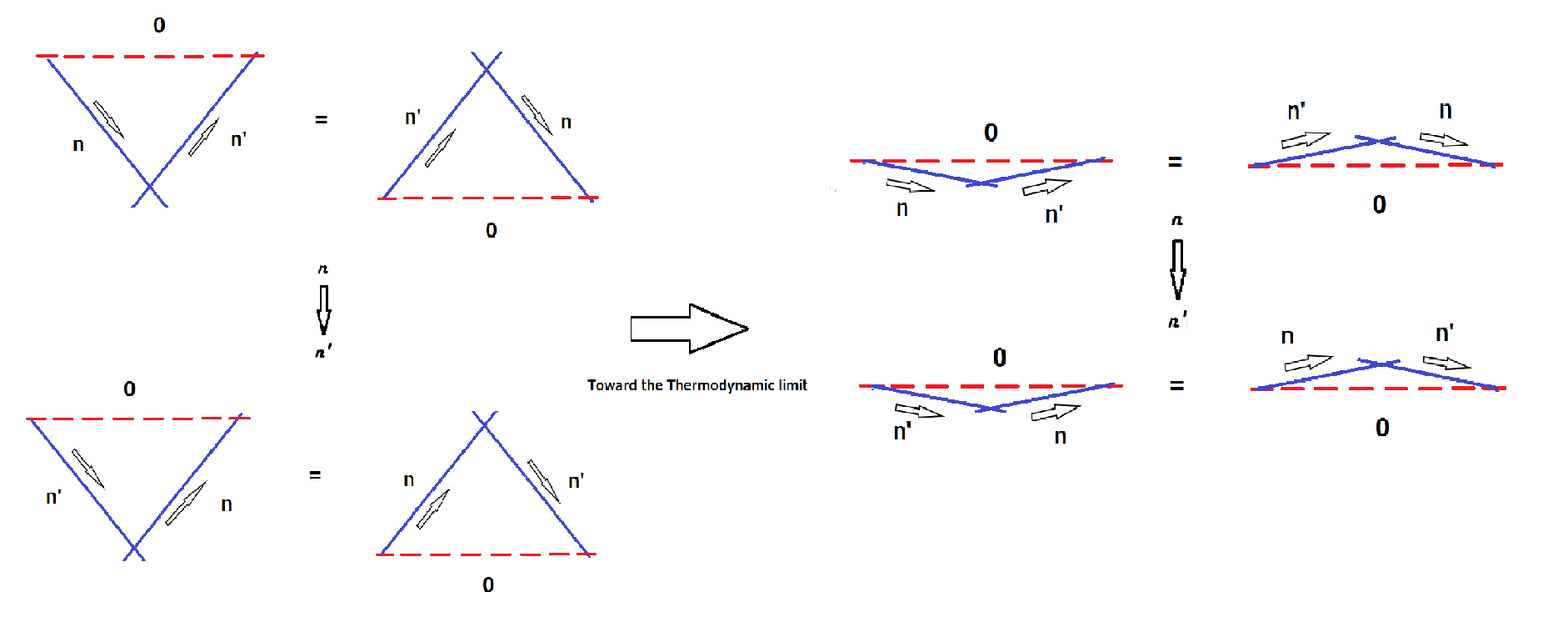}
	\caption{The triangular representation of the vacuum degeneracy. At the finite volume approximation, the triangles are large (left hand-side). At the infinite volume limit, the triangles become narrow and the three lines (in the triangle) become almost parallel (right-hand side of the figure). The equality between triangles in the figure represents the QYBE and each triangle corresponds to one term in the expansion (\ref{nojoda}) after summing over the degenerate vacuum.}
	\label{Iwillbeback2}
\end{figure}

\begin{figure}
	\centering
		\includegraphics[width=14cm, height=8cm]{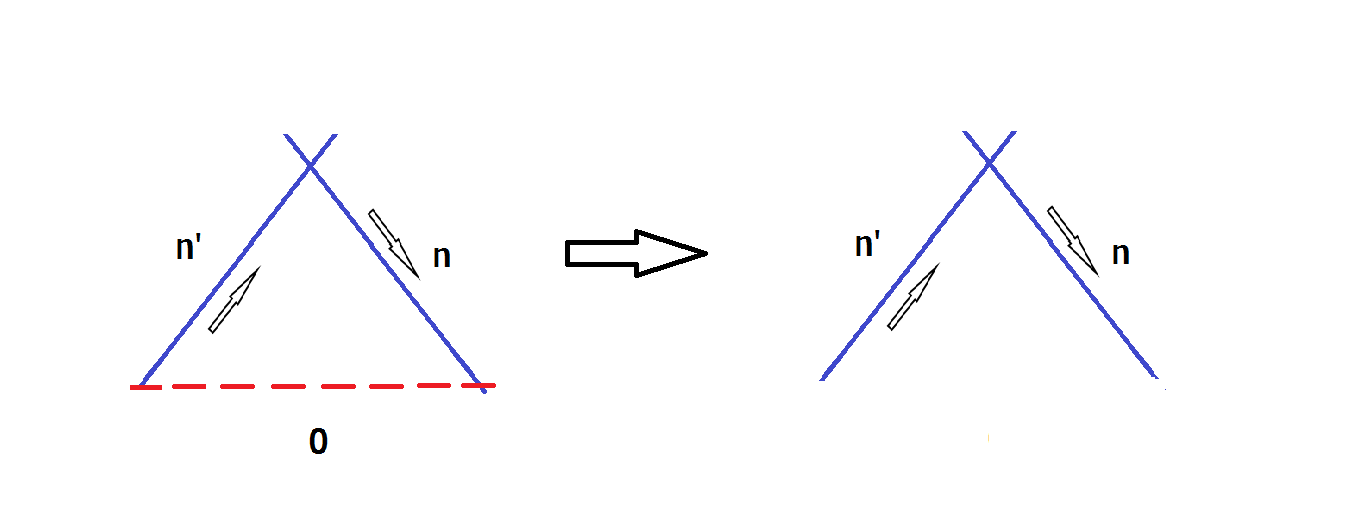}
	\caption{The effect of external perturbations. In the infinite volume approximation, any small external perturbation selects only one vacuum arbitrarily. The left-hand side represents the situation before the perturbation appears. In such a case, the QYBE are valid and we can sum over the degenerate vacuum. Once the perturbation selects one particular vacuum, the degenerate vacuum disappears (red dotted line) and we have the usual interaction of two particles (two Nambu-Goldstone bosons). The Nambu-Goldstone bosons however, only move through the degenerate vacuum. The slopes of the lines are exaggerated in the figure but in the large volume limit the lines are almost horizontal.}
	\label{Iwillbeback21}
\end{figure}

In this section we have explained that summing the vacuum expectation value of the order parameter over the degenerate vacuum gives a trivial result defined in eq. (\ref{here la}). This justifies the step going from eq. (\ref{nojoda}) to eq. (\ref{nojoda3}) and it is a natural consequence of the QYBE. In the thermodynamic limit, the different vacuums are orthogonal because they belong to different Fock spaces (Hilbert spaces). If we have a coherent vacuum state for example, the broken generators cannot annihilate the vacuum but they rather rotate it toward a new one in the following form

\begin{equation}
\vert\theta>_0\to U_{\theta'}\vert\theta>_0=\vert\theta+\theta'>.
\end{equation}  
Here $U_\theta=e^{i\theta Q}$, with $Q$ representing the broken generator \cite{Tomas4}. It can be demonstrated that in the thermodynamic limit

\begin{equation}   \label{otho}
\vert<\theta'\vert\theta>_0\vert\to0.
\end{equation}
Then the vectors contained in the Fock space expanded around the vacuum $\vert\theta>_0$ are orthogonal to the vectors in the Fock space expanded by the vacuum $\vert\theta'>_0$. Then the Hilbert spaces involved are different in principle. The reality is that the symmetry is not spontaneously broken at finite volume approximation and in reality the vacuums are never perfectly degenerate. In this limit, the Nambu-Goldstone bosons can move from one vacuum to the next after tunneling through some potential barriers \cite{Tomas4}. The vacuums are almost degenerate at the infinite volume limit and in such a case the Nambu-Goldstone bosons can move freely from one vacuum to the next as it is remarked in \cite{Tomas, Tomas4}. In this limit any small perturbation selects one particular vacuum state. The calculations done in this paper are valid in general. This is the case because once we take the expectation value of the order parameter over some specific vacuum, what we have is a number. Evaluating over other vacuums, will give us a complex number in general. Nothing forbids us to sum complex numbers. However, it is interesting to analyze the physics behind the QYBE. In this sense, the Yang-Baxter diagrams for modeling the spontaneous symmetry breaking phenomena, can be considered to be valid either, at the finite volume approximation where a single Hilbert space can be assumed, after which at the end we can take the thermodynamic limit. Or we can consider the diagrams to be valid in the Thermodynamic limit but under the ideal absence of external perturbations. This is just an idealization and the Nambu-Goldstone bosons keep memory of this statement. In \cite{Tomas}, it has been calculated the exact product between vacuums to be

\begin{equation}
\vert<\theta'\vert\theta>_0\vert=e^{-\Omega\vert\theta-\theta'\vert^2},
\end{equation}       
where $\Omega$ is just the volume of the system. This confirms that at finite volume approximation, the vacuums are non-orthogonal. In addition, in the same limit the momentum of the intermediate particles is non-zero as can be observed from the illustration given in the figure \ref{Iwillbeback2}. Note that on the left-hand side of the Fig. \ref{Iwillbeback2} the slopes of the lines representing the Nambu-Goldstone bosons $n$ and $n'$ are larger than the slopes taken at the thermodynamic limit (right-hand side of the same figure). In the thermodynamic limit, the three lines in the triangle are (almost) parallel and the Nambu-Goldstone bosons are gap-less particles. If there is any external fluctuation at the thermodynamic limit, then the degeneracy of the vacuum is reduced to only one arbitrarily selected vacuum as can be observed from the figure \ref{Iwillbeback21}. In the same figure, when the red-dotted line disappears, then there is no degeneracy of the vacuum. Independent of the situation under analysis, nothing forbids us in summing over the degenerate vacuum and this can be interpreted as a superposition of the different vacuum expectations value of the order parameter which gives the trivial result (\ref{here la}). The result has a direct correspondence with the QYBE. It is important to remark that the spontaneous symmetry breaking condition is time-independent only in the thermodynamic limit \cite{Tomas}.   

\subsection{The theorems contained inside the Yang-Baxter relations}

There are a series of theorems and rules already established around the dispersion relations as well as on the number of Nambu-Goldstone bosons appearing in a system \cite{Murayama, Tomas}. In this subsection we want to remark some rules observed from the QYBE when they are applied to the spontaneous symmetry breaking situation.     
First, let's note that eq. (\ref{here la}) means that if we sum over the degenerate vacuum the result (\ref{impo}), then we obtain

\begin{equation}   \label{impo55}
\sum_0<0_{DV}\vert \left[\phi_{b,a}(x), [Q_{p,k}(y),Q_{l,m}(z)]\right]\vert0_{DV}>=0.
\end{equation}
This expression, contains two different QYBE given by eqns. (\ref{eq:the last option2}) and (\ref{eq:the last optionmiau}). In fact, if we expand the previous result, we can express eq. (\ref{nojoda}) in terms of the $R$-matrices after summing over all the possible vacuums. We can then obtain the result

\begin{eqnarray}   \label{nojodadelajoda}
R^{n',0}_{l,m}R^{k,n}_{n',p}R^{b,a}_{0,n}e^{i(\tilde{p}^0_{n'}z^0+{\bf\tilde{p}}_{n'}\cdot {\bf z})}-R^{n,0}_{p,m}R^{a,n'}_{n,l}R^{b,k}_{0,n'}e^{i(p^0_{n}y^0+{\bf p}\cdot{\bf y})}\nonumber\\
-R^{0,n}_{m,p}R^{n',a}_{l,n}R^{k,b}_{n',0}e^{i(p^0_{n}y^0-{\bf p}\cdot{\bf y})}+R^{0,n'}_{m,l}R^{n,k}_{p,n'}R^{a,b}_{n,0}e^{i(\tilde{p}^0_{n'}z^0-{\bf\tilde{p}}_{n'}\cdot {\bf z})}=0.
\end{eqnarray}
Here we have omitted the trivial phases containing differences in the four-momentum ($p_n-p_{n'}$) and similar terms. In the infinite volume limit, they just represent energy-momentum conservation. Note in addition that in eq. (\ref{nojodadelajoda}), the phases corresponding to the first two terms in the expansion, appear in a time-reversed version. The reason for this comes out naturally from the QYBE and it is determined by the order how the $R$-matrices appear. Note that we are considering 

\begin{equation}   \label{ahayque}
R^{n',0}_{l,m}R^{k,n}_{n',p}R^{b,a}_{0,n}=\sum_{0, n, n'}<0_{DV}\vert \phi_{b}\vert n><n\vert Q_{p}(0)\vert n'><n'\vert Q_{l}(0)\vert0_{DV}>,
\end{equation}
\begin{equation}   \label{ahayque2}
R^{n,0}_{p,m}R^{a,n'}_{n,l}R^{b,k}_{0,n'}=\sum_{0, n, n'}<0_{DV}\vert \phi_{b}\vert n'><n'\vert Q_{l}\vert n><n\vert(0)Q_{p}(0)\vert0_{DV}>.
\end{equation}
Since the order of the $R$-matrices is reversed with respect to the operators which they represent, then it is necessary to make the corresponding corrections to the phases in eq. (\ref{nojodadelajoda}). Note that the product of the $R$-matrices and their equivalence with respect to the product of operators, is determined by the spaces defined by the lines in the triangular relations in the QYBE. In simple words, the index contractions determines the correspondence between the $R$-matrices products and the product of operators. We can then conclude that the triangles corresponding to the previous terms, determine the appropriate phases naturally. This is a remarkable property of the method presented in this paper. Due to the QYBE defined in eqns. (\ref{eq:the last option2}) and (\ref{eq:the last optionmiau}), we only have two independent terms in eq. (\ref{nojodadelajoda}) and we can reduce eq. (\ref{nojodadelajoda}) to  

\begin{eqnarray}   \label{nojodadelajodadelajoda}
R^{n',0}_{l,m}R^{k,n}_{n',p}R^{b,a}_{0,n}\left(e^{i(\tilde{p}^0_{n'}z^0+{\bf\tilde{p}}_{n'}\cdot {\bf z})}-e^{i(p^0_{n}y^0-{\bf p}\cdot{\bf y})}\right)\nonumber\\
-R^{n,0}_{p,m}R^{a,n'}_{n,l}R^{b,k}_{0,n'}\left(e^{i(p^0_{n}y^0+{\bf p}\cdot{\bf y})}-e^{i(\tilde{p}^0_{n'}z^0-{\bf\tilde{p}}_{n'}\cdot {\bf z})}\right)=0.
\end{eqnarray}
Having this expression, we can define the following theorems:\\
{\bf Theorem 1}: If we have two broken generators in a system, which together with Nambu-Goldstone field, satisfy the condition $R^{n',0}_{l,m}R^{k,n}_{n',p}R^{b,a}_{0,n}=R^{n,0}_{p,m}R^{a,n'}_{n,l}R^{b,k}_{0,n'}$, under the definitions given in eq. (\ref{ahayque}) and (\ref{ahayque2}), then there is a quadratic dispersion relation for the Nambu-Goldstone bosons ($E_{\bf p}\backsim {\bf p}^2$) and there is one Nambu-Goldstone boson corresponding to two broken symmetries.\\\\
Note that in this theorem, the equality is done over the $R$-matrices, but it can be also expressed as

\begin{eqnarray}
\sum_{0, n, n'}<0_{DV}\vert \phi_{b}\vert n><n\vert Q_{p}(0)\vert n'><n'\vert Q_{l}(0)\vert0_{DV}>=\nonumber\\
\sum_{0, n, n'}<0_{DV}\vert \phi_{b}\vert n'><n'\vert Q_{l}\vert n><n\vert(0)Q_{p}(0)\vert0_{DV}>,
\end{eqnarray}
in agreement with the expressions (\ref{ahayque}) and (\ref{ahayque2}). The theorem suggests clearly that under the consecutive action over the different vacuum states of the Nambu-Goldstone field plus a pair of broken generators; if the order (of action) of the broken generators is not important, then both generators are redundant and they both correspond to a single degree of freedom. We can also perceive this with the statement that we only have one independent history of interaction in this case. This statement is consistent with the results found in \cite{Murayama, Tomas}. The second theorem to be formulated is just a corollary of the first one. \\
{\bf Corollary}: If we have two broken generators in a system, which together with the Nambu-Goldstone field, satisfy the inequality condition $R^{n',0}_{l,m}R^{k,n}_{n',p}R^{b,a}_{0,n}\neq R^{n,0}_{p,m}R^{a,n'}_{n,l}R^{b,k}_{0,n'}$, under the definitions given in eqns. (\ref{ahayque}) and (\ref{ahayque2}), then there is a linear dispersion relation for the Nambu-Goldstone bosons ($E_{\bf p}\backsim {\bf p}$) and there is one Nambu-Goldstone boson for each broken generator.\\
This Corollary can be explained in the following way. Under the previous inequality condition, the pair of terms appearing in the expansion (\ref{nojodadelajodadelajoda}) are independent and then each term must vanish separately. This means that we have two independent conditions defined as

\begin{eqnarray}   \label{gaga}
R^{n',0}_{l,m}R^{k,n}_{n',p}R^{b,a}_{0,n}\left(e^{i(\tilde{p}^0_{n'}z^0+{\bf\tilde{p}}_{n'}\cdot {\bf z})}-e^{i(p^0_{n}y^0-{\bf p}\cdot{\bf y})}\right)=0,\nonumber\\
R^{n,0}_{p,m}R^{a,n'}_{n,l}R^{b,k}_{0,n'}\left(e^{i(p^0_{n}y^0+{\bf p}\cdot{\bf y})}-e^{i(\tilde{p}^0_{n'}z^0-{\bf\tilde{p}}_{n'}\cdot {\bf z})}\right)=0.
\end{eqnarray}
Here we have naturally a linear dispersion relation if we evaluate the limit when the momentum and energy go to zero simultaneously. This analysis is consistent with what we have described before in eq. (\ref{nojodadelajodajoda22}). The result (\ref{gaga}) means that under the consecutive action, over the vacuum states, of a pair of broken generators plus the Nambu-Goldstone field; if the order of operation of the pair of broken generators is important, then they are not redundant and each broken generator is related to a single degree of freedom (Nambu-Goldstone boson). In \cite{My papers} some preliminary and complementary analysis with respect to the one developed in this paper were done. However, inside the knowledge of the author, this is the first time that theorems related to the dispersion relation together with the number of Nambu-Goldstone bosons are done by using the QYBE.   

\section{Conclusions}   \label{Conclusions}

In this paper we have derived a novel method for understanding the interaction, counting, as well as the dispersion relations for the Nambu-Goldstone bosons in general. When the symmetry under exchange $n\to n'$ is satisfied ($n=n'$), then the eqns. (\ref{eq:the last option2}) and (\ref{eq:the last optionmiau}) are equivalent. Then a quadratic dispersion relation will appear for the associated Goldstone bosons after including the corresponding phases. The QYBE are then consistent with the the interactions of Goldstone bosons and they can be interpreted with the fact that under some circumstances, two Goldstone bosons with linear dispersion relation interact in order to produce effectively a single one with quadratic dispersion relation. On the other hand, if $n\neq n'$, then eqns. (\ref{eq:the last option2}) and (\ref{eq:the last optionmiau}) turn out to be different and then we will obtain a linear dispersion relation. In addition, the rank of the $R$-matrices appearing in the QYBE is related to the number of Nambu-Goldstone bosons and the number of broken generators. This is the case because the $R$-matrices represent either, the interaction between pairs of Goldstone bosons or the interaction between Goldstone bosons and the degenerate vacuum. The result is summarized in eq. (\ref{counting}). In this paper we could formulate a theorem and a corollary connecting the $R$-matrices appearing in the QYBE with the observed dispersion relations. The theorem simultaneously connects the number of broken generators with the number of Nambu-Goldstone bosons. The implementation of the QYBE requires a sum over the degenerate vacuum which is equivalent to a superposition of the vacuum expectation values of the order parameter evaluated over all the possible (degenerate) vacuums. This sum is evidently zero. We have made the corresponding Mathematical justification for summing over the degenerate vacuum the different expressions connected to the spontaneous symmetry breaking phenomena. The results show the consistency of the QYBE with the symmetry breaking mechanism. Future interesting applications of this method will be explored in coming papers. Before Watanabe, Brauner and Murayama worked over some interesting approach to the problem of counting Nambu-Goldstone bosons as well as explaining the observed dispersion relations \cite{Murayama}. The approach in this paper is different because it is based on the dynamical constraints emerging from the QYBE. In \cite{Murayama1, Murayama2, Murayama}, although the authors did remarkable contributions to the problem, it was never mentioned by then the important role of the QYBE in the solution of this important problem. \\

{\bf Acknowledgement}
I thank Michio Jimbo for wonderful discussions about the Quantum Yang-Baxter equations at Rikkyo University, Ikebukuro campus. I thank E. Witten as well as J. Maldacena for wonderful lectures during the conference "Strings 2016" and for wonderful discussions after. The lecture of E. Witten was fundamental for the formulation of this idea which started to be done during the conference Strings 2016 organized in Beijing at Tsinghua University. I thank Tadashi Takayanagi for his kind attention during my visit to the Yukawa Institute for Theoretical Physics (YITP). I thank Takahiro Nishinaka for nice discussions about these results during my visit to YITP hosted by Tadashi Takayanagi. I thank Jan de Boer for his wonderful comments and questions during my presentation in the NCTS Annual Theory Meeting, organized at the National Tsinghua University in Hsinchu, Taiwan. Finally, I thank Benjamin Basso for interesting questions and comments. This work was partially supported by the JSPS Post-Doctoral fellow for oversea researchers and for the Open University of Hong Kong.                

\bibliography{basename of .bib file}

\end{document}